\documentclass[fleqn,usenatbib]{mnras}

\usepackage{newtxtext,newtxmath}

\usepackage[T1]{fontenc}

\DeclareRobustCommand{\VAN}[3]{#2}
\let\VANthebibliography\thebibliography
\def\thebibliography{\DeclareRobustCommand{\VAN}[3]{##3}\VANthebibliography}

\usepackage{graphicx}	
\usepackage{amsmath}	
\usepackage{pdflscape}   
\usepackage{tabularx}    
\usepackage{booktabs} 

\title[Spectroscopy of brown dwarfs in IC 1396]{Revealing the substellar population of IC 1396: A spectroscopic survey of brown dwarfs in the region}

\author[B. Damian et al.]{
Belinda Damian$^{1,2}$\thanks{bd64@st-andrews.ac.uk (BD)},
Jessy Jose$^{3}$\thanks{jessyvjose1@gmail.com (JJ)},
Aleks Scholz$^{1}$,
Carlos Román Zúñiga$^{4}$,
B. Shridharan$^{5}$,
\newauthor
Genaro Suárez$^{6}$,
Juan José Downes$^{7}$,
Swagat R. Das$^{8}$, 
Sreeja S. Kartha$^{2}$
and Saumya Gupta$^{9}$
\\\\
$^{1}$SUPA, School of Physics \& Astronomy, University of St Andrews, North Haugh, St Andrews, KY169SS, United Kingdom\\
$^{2}$Department of Physics and Electronics, CHRIST (Deemed to be University), Hosur Road, Bengaluru 560029, India\\
$^{3}$Department of Physics, Indian Institute of Science Education and Research (IISER) Tirupati, Yerpedu, Tirupati, Andhra Pradesh, 517619, India\\
$^{4}$Universidad Nacional Autónoma de México, Instituto de Astronomía, AP 106, Ensenada 22800, BC, Mexico\\
$^{5}$Department of Astronomy and Astrophysics, Tata Institute of Fundamental Research, Homi Bhabha Road, Colaba, Mumbai 400005, India\\
$^{6}$Department of Astrophysics, American Museum of Natural History, Central Park West at 79th Street, NY 10024, USA\\
$^{7}$Departamento de Astronomía, Facultad de Ciencias, Universidad de la República, Iguá 4225, Montevideo, CP 11400, Uruguay\\
$^{8}$Departamento de Astronom{\'i}a, Universidad de Chile, Las Condes, 7591245 Santiago, Chile\\
$^{9}$Astronomy Unit, School of Physics and Astronomy, Queen Mary University of London, London E1 4NS, UK
}

\date{Accepted XXX. Received YYY; in original form ZZZ}

\pubyear{2025}

\begin{document}
\label{firstpage}
\pagerange{\pageref{firstpage}--\pageref{lastpage}}
\maketitle

\begin{abstract}
We present a new spectroscopic view of the brown dwarf population in the young star-forming region IC 1396 and investigate the impact of environment on low-mass star formation. We use deep optical photometry from Subaru-HSC to identify the candidate low-mass stars and brown dwarfs in the region. Our follow-up low-resolution spectroscopic survey with GTC-EMIR and IRTF-SpeX has identified 32 new members in the region with spectral types between M3 and M9, among which 25 are brown dwarfs with spectral types M6 or later. We use the BT-Settl atmospheric models to derive the effective temperatures of the members. Using a comprehensive catalogue of known members and candidates, we estimate the star to brown dwarf ratio for IC 1396 to be 5.0$\pm$0.4, for a mass range between 1-0.03 M$_\odot$. This ratio is largely consistent with measurements in other young clusters spanning a range of UV radiation fields and stellar densities, supporting formation scenarios in which the relative abundance of brown dwarfs is not strongly influenced by the local environmental conditions. 
\end{abstract}




\section{Introduction}
\label{sec:intro}
Multiple mechanisms have been proposed for the formation of brown dwarfs, objects with masses below the hydrogen burning limit of  $\sim$0.075 M$_\odot$ and extending down to $\sim$0.012 M$_\odot$. One possibility is that they form through the same process as stars, through the collapse and fragmentation of the molecular cloud cores.  Over the past two decades, numerous observational studies have shown that brown dwarfs share similarities with low-mass stars (see review by \citealt{luhman2012}). Key observational properties, for example the mass function, the binarity fraction, or the disk mass, change smoothly as a function of the object mass, suggesting that brown dwarfs and low-mass stars may form through related physical processes (e.g., \citealt{scholz2012,damian2023a,muzic2025}). However there are also alternate scenarios that explain the formation of these objects like ejection of the molecular embryos from multiple systems, gravitational instability of the protoplanetary disks and photo-erosion of prestellar cores in the vicinity of massive stars (\citealt{bate2012,offner2014}). The relative contribution of these different mechanisms, and the impact of feedback from the stellar environment in the formation of brown dwarfs has not been fully understood.

Theoretical studies predict that environmental conditions can alter the brown dwarf formation efficiency. For example, simulations involving dynamical interactions and ejection scenarios in dense cluster environments \citep{bate2012}, as well as gravitational fragmentation of infalling gas \citep{bonnell2008}, predict that higher stellar densities can enhance brown dwarf formation, with an order of magnitude increase in density potentially resulting in a factor of two increase in the brown dwarf fraction. Apart from stellar density, photo-erosion models suggest that UV radiation from nearby OB stars may also influence brown dwarf formation by truncating accretion onto low-mass prestellar cores \citep{whitworth2004}. The path to observationally test such scenarios is to conduct deep surveys for brown dwarfs in young star-forming environments spanning a range of initial conditions. Taken together these studies will ultimately be able to establish brown dwarf numbers as a function of cluster density, radiation environment, or evolutionary stage (\citealt{abad2023,gupta2024,muzic2025}). Most existing brown dwarf surveys have primarily targeted the nearest star-forming regions, which are typically low density environments that lack massive OB stars, with the notable exception of the Orion Nebula Cluster. Over the past years, several groups have used the largest ground-based telescopes and JWST to extend these surveys to larger distances and thereby probe more diverse regions (\citealt{muzic2017,jose2020,abad2023,defurio2025}). Typically, searches for young brown dwarfs begin with a photometric survey, followed by spectroscopy of selected candidates that exhibit the characteristic colours of young substellar objects (\citealt{abad2023, damian2023a, luhman2024}).

\begin{figure}
    \centering
    \includegraphics[width=\linewidth]{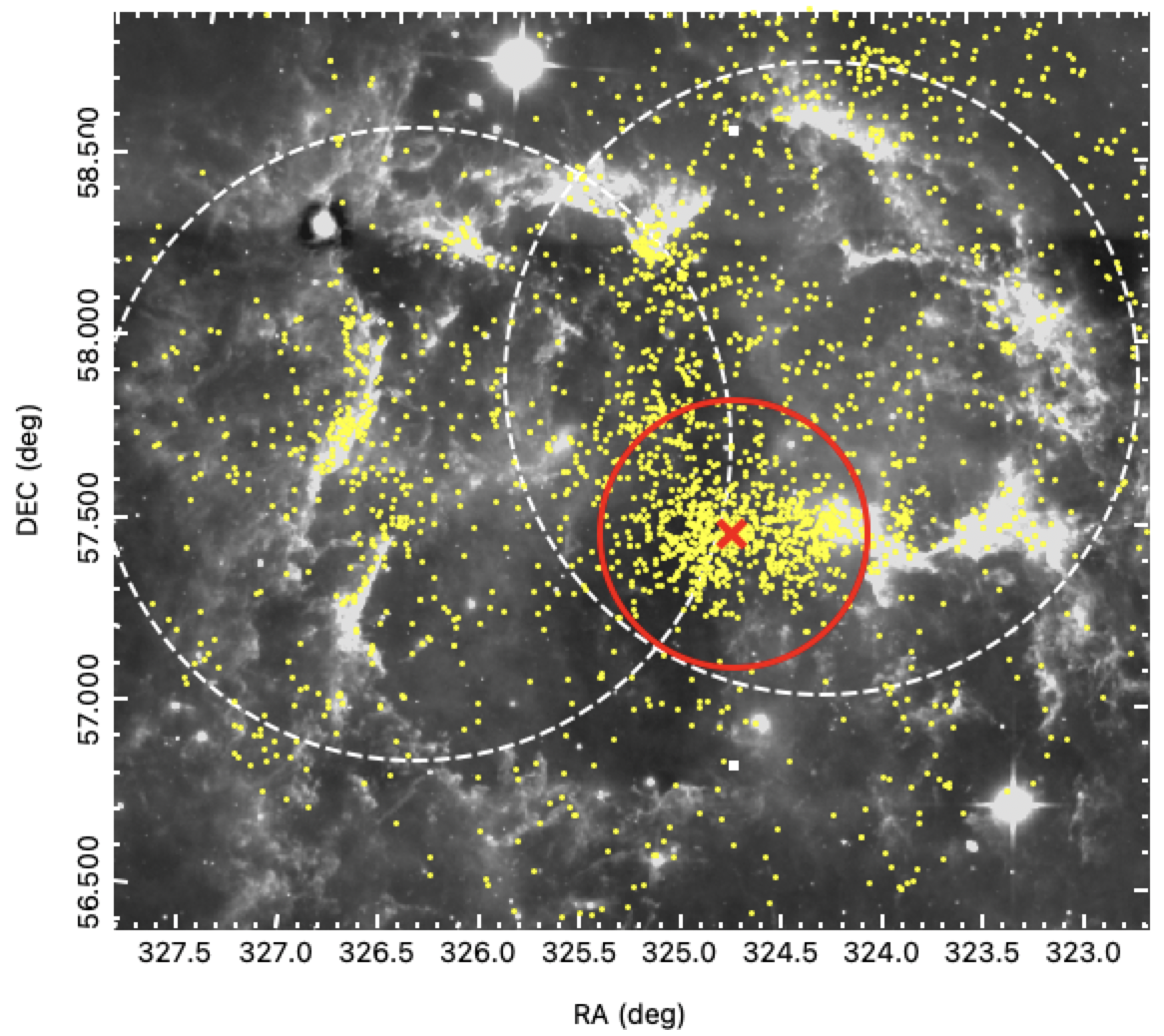}
    \caption{WISE 22 $\mathrm{\mu}$m image of the IC 1396 complex. The white dashed circles represent the two pointings of the Subaru-HSC observations. The red circle marks the area analysed in this study, with a radius of 22$\arcmin$ ($\sim$6 pc), centred on the massive star HD 206267, shown by the red cross. Previously confirmed members and candidates are marked with the yellow dots.}
    \label{fig:ic1396_wiseimage}
\end{figure}

This paper presents a continuation of these efforts by carrying out a spectroscopic survey of substellar objects in the young, massive star-forming region IC 1396. IC 1396 is a young star-forming HII region in the Cepheus OB2 association, spanning nearly 3$^{\circ}$ in diameter \citep{das2025} at a distance of $\sim$915-945 pc \citep{sicilia2019,das2023}. This region, with an age of 2-4 Myr \citep{sicilia2005,das2023,pelayo2023} is irradiated by a massive multiple system HD 206267 (O5+O9) \citep{pelayo2023} located at the centre of the cluster. The intense radiation and stellar winds from these massive stars, have cleared out the surrounding gas and dust creating a noticeable cavity at the centre with relatively low extinction (A$_\mathrm{V}$ $\sim$1 - 2 mag; \citealt{nakano2012,sicilia2013}). Additionally, the region is associated with several bright-rimmed clouds (BRCs), globules, and the prominent elephant-trunk structure (IC 1396A) \citep{froebrich2005}. The young age, uniform low reddening, and intense feedback-driven environment make IC 1396 one of the best regions beyond the solar neighbourhood to detect brown dwarfs and conduct in-depth investigations into the factors influencing it. Figure~\ref{fig:ic1396_wiseimage} shows a 22$\mu$m Wide-field Infrared Survey Explorer (WISE) image of the IC 1396 complex and highlights the area targeted in this study.

A range of studies have identified members in this region using various approaches, including photometric and spectroscopic surveys \citep{contreras2002,barentsen2011,sicilia2013}, X-ray data \citep{getman2012}, infrared disk indicators \citep{morales2009,silverberg2021}, and Gaia-based membership \citep{das2023,pelayo2023}. More recently, \citet{gupta2024} and \citet{das2025} identified low-mass candidates using machine learning techniques harnessing deep multi-band imaging of the central region (radius $\sim$22$\arcmin$) around the massive star and the entire complex (radius $\sim$1.5$^\circ$), respectively. Prior to these studies, however, there was neither a comprehensive photometric survey of the substellar population nor any spectroscopic confirmation of brown dwarfs in this region. In this work, we concentrate on a 22$\arcmin$ radius around the central massive star HD 206267 (RA: 324.74$^{\circ}$; Dec: 57.49$^{\circ}$), where over 60\% of the previously known cluster members are present \citep{gupta2024}.

This paper is structured as follows: in Section~\ref{sec:data} we discuss the optical photometric catalogue that is used and the near-infrared (NIR) spectroscopic observations and data reduction steps. In Section~\ref{sec:target_BDcharact} we present the selection of candidate substellar objects for follow-up spectroscopy and the spectral characterisation of the targets. Section~\ref{sec:discussion} presents the results and discussion and Section~\ref{sec:summary} gives the summary of the paper.

\section{Observations and data reduction}
\label{sec:data}
The optical photometric catalogue from \citet{das2025} is used to identify the candidate brown dwarfs in the region. A summary of this data is given in Section~\ref{sec:data_photo}. For details on the data reduction technique and catalogue compilation refer to \citet{das2025} and \citet{gupta2024}. In Section~\ref{sec:data_spec}, we discuss the follow-up spectroscopy of the candidate brown dwarfs in the IC 1396 cluster with two spectrographs - EMIR in GTC and SpeX in IRTF. 

\subsection{Optical photometry}
\label{sec:data_photo}
\subsubsection{Subaru-HSC}
The optical data presented in \citet{das2025} was obtained from the Hyper Suprime-Cam (HSC) instrument on the 8.2 m Subaru Telescope. These observations were carried out on 18 September 2017 (PI: Jessy Jose, Program ID: S17B0108N) for two pointings, each covering an area of 1.5$^{\circ}$ diameter, with central coordinates at RA: 324.27$^{\circ}$; Dec: 57.91$^{\circ}$ and RA: 326.37$^{\circ}$; Dec: 57.72$^{\circ}$ with 
three broad-band optical filters r$_2$, i$_2$, and Y\footnote{Central wavelengths of r$_2$, i$_2$, and Y bands are 622.6 nm, 776.7 nm, and 1005.1 nm respectively \citep{kawanomoto2018}.}. The data was processed using the HSC pipeline version 6.7 following the procedures outlined in \citet{gupta2021}.

Within the area focussed in this study (i.e. 22$\arcmin$ radius around the ionizing source HD 206267), there are $\sim$126600 sources detected, with photometry in all the three bands and photometric errors $<$0.2 mag. We estimate the 90\% completeness limit using the turnover point of source count approach (\citealt{damian2021,damian2024}) to be 24.5 mag in the r$_2$ band, 22.5 mag in the i$_2$ band, and 21.5 mag in the Y band. According to the \citet{baraffe2015} models, this corresponds to $\sim$0.03 M$_\odot$, at a distance of 917 pc for an extinction (A$_\mathrm{V}$) of 1 mag and an age of 2 Myr (see Section~\ref{sec:target_selection}).

\subsubsection{Pan-STARRS}
In addition to the HSC data, Pan-STARRS DR1 data  \citep{chambers2016} in r$_{P1}$, i$_{P1}$, and Y$_{P1}$ \footnote{Effective wavelengths of r$_{P1}$, i$_{P1}$, and y$_{P1}$ bands are 617 nm, 752 nm, and 962 nm respectively \citep{tonry2012}.} filters was used to construct the optical catalogue in \citet{das2025}. Sources with photometric errors in individual filters $<$0.2 mag were selected. Transformation equations from \citet{gupta2021} were applied to convert photometry from the Pan-STARRS filter system to the HSC system. The Pan-STARRS photometry primarily supplements the HSC photometry by introducing sources, particularly at the bright end, that may have been missed due to saturation. Henceforth, for simplicity we refer to this combined optical catalogue as the HSC catalogue.

\subsection{Spectroscopic follow-up}
\label{sec:data_spec}
\subsubsection{GTC-EMIR}
A spectroscopic survey of a subset of our brown dwarf candidates (see Section~\ref{sec:target_selection} for details on target selection) was done with the EMIR (Espectrógrafo Multiobjeto Infra-Rojo) instrument on the 10.4 m Gran Telescopio CANARIAS (GTC), situated in the Roque de los Muchachos Observatory at La Palma, Canary Islands. EMIR is a NIR wide-field imager and a medium resolution multi-object spectrograph with a field of view of 4.0$^\prime$$\times$6.67$^\prime$ for spectroscopy \citep{garzon2016}. We employed EMIR in Multi-object Spectroscopy (MOS) mode to observe the candidate brown dwarfs in October 2022 (PI: Carlos R. Zuniga, Program ID: GTC4-22BMEX). The observations were conducted in low-resolution HK mode (R$\sim$900) covering a wavelength range of 1.45-2.42 $\mu$m. Twenty nine targets were observed with four mask positions designed to accommodate the targets, with the number of slitlets optimised for observations in the ABBA mode. Additionally, each mask included three isolated bright sources as reference stars to facilitate the pointing. The targets were observed in four observing blocks, one for each mask position, with two separate observing blocks for the standard star (HD 199217) interspersed between the targets blocks. Each target and standard star observing block consisted of 5 and 4 ABBA cycles with a single exposure time of 60 sec and 5 sec, respectively.

The raw data was reduced using the dedicated EMIR data reduction pipeline, PyEmir (version 0.19) \citep{pascual2010}. The reduction was performed following the procedure given in the PyEmir tutorial\footnote{\url{https://guaix-ucm.github.io/pyemir-tutorials/index.html}} and the steps are briefly summarised here. We used the standard pipeline steps to rectify and wavelength calibrate the raw images of the standard star and the targets. To calibrate the flux of the sources, we first extracted the spectra of each target in the calibrated image dispersed along the respective slit positions. We then used the atmospheric model of an A0 standard star from \citet{castelli2003}. This model spectra was scaled by the (R/D)$^2$ factor, which was obtained using the synthetic photometry of our standard star (HD 199217) and by comparing the SED of this star with the \citet{castelli2003} model. This estimate was done using the online tool, \texttt{VOSA} SED analyser\footnote{\url{http://svo2.cab.inta-csic.es/theory/vosa/}}. The observed spectra of HD 199217 was divided by the scaled model spectra to obtain a preliminary response curve and it was corrected for telluric lines. To check the accuracy of the calibration process we deduced the synthetic H and K band photometry from the calibrated spectra of HD 199217 by convolving the flux for the 2MASS filter profile and found the difference to be $<$0.1 mag with the 2MASS photometry. We then use the response curve of the standard star to flux calibrate all the targets. Out of the 29 targets that were observed, the spectra of 12 of them were extracted with sufficient signal-to-noise ratio and analysed further as discussed in Section~\ref{sec:target_BDcharact}.

\subsubsection{IRTF-SpeX}
Additionally, we carried out spectroscopy with the SpeX instrument on the 3.2 m NASA Infrared Telescope Facility (IRTF) \citep{rayner2003} for 33 candidate low-mass stars and brown dwarfs (three targets observed with EMIR were also observed with SpeX) in low-resolution prism mode (R$\sim$150). The observations were conducted in September 2021 and July 2022 (PI: Belinda Damian, Program IDs: 2021B041, 2022A056). SpeX in low-resolution mode covers a spectral range of 0.7-2.5 $\mu$m, which is ideal for studying the water absorption and gravity-sensitive features in the NIR regime \citep{allers2013}.

Depending on the seeing conditions, we used the slits of width 0.5$^{\prime\prime}$ or 0.8$^{\prime\prime}$. For each observation, we employed the standard ABBA nodding pattern to capture both the sky and target spectra. The integration time for each exposure (120-180 sec) and the number of ABBA cycles (4-6) varied depending on the brightness of the source.  To ensure accurate wavelength calibration, flat field and argon lamp observations were taken between the target observations. Additionally, we observed a nearby A0V standard star every hour to perform telluric correction and flux calibration. 

The data reduction process was carried out using the IDL based spectral reduction tool, \texttt{Spextool} (version 4.1) \citep{cushing2004} following the procedure detailed in \citet{jose2020}. The individual spectra from the set of target observations were extracted and wavelength calibrated. To mitigate the impact of bad pixels on flux variability, all the spectra in the sequence for a given target were scaled to a common median flux and then median combined using the \texttt{xcombspec} GUI. Subsequently, to correct for telluric absorption and flux calibration of the combined target spectra, we utilized the standard star spectra and a high-resolution model of Vega in the xtellcor GUI \citep{vacca2003}. Finally, we extracted the spectra of 30 sources (one of which was also observed with GTC-EMIR) with good signal-to-noise ratio, sufficient to characterise the objects.

\section{Target selection and spectral characterisation}
\label{sec:target_BDcharact}
\subsection{Candidate selection for spectroscopy}
\label{sec:target_selection}
\begin{figure}
    \centering
    \includegraphics[scale=1.2]{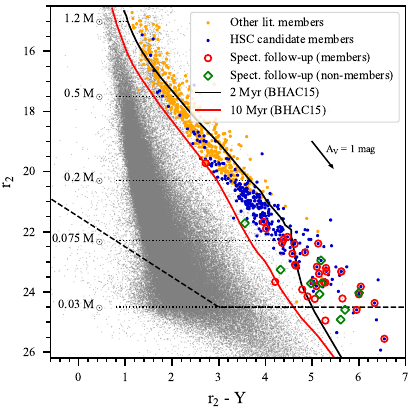}
    \caption{Optical CMD with HSC photometry. The grey dots show the sources detected in the survey area of 22$\arcmin$ radius (refer Section~\ref{sec:data_photo}). From \citet{das2025} we highlight their candidates in blue and the Gaia/non-Gaia based members identified in literature in orange (refer text for details). The 41 sources with spectra obtained in this work are marked with red open circles (members) and green open diamonds (non-members). The black dashed line marks the 90\% completeness limit of our optical catalogue. The black and red solid lines are the 2 Myr and 10 Myr \citet{baraffe2015} isochrones, respectively,  reddened by A$_\mathrm{V}$ = 1 mag. The masses corresponding to the 2 Myr isochrone are also shown with the dotted lines.} 
    \label{fig:cmd2}
\end{figure}

The candidate low-mass stars and brown dwarfs for the spectroscopic survey, were selected using an optical colour-magnitude diagram (CMD) with the HSC data as shown in Figure~\ref{fig:cmd2}.The figure shows the cluster members compiled in \citet{das2025} drawn from several previous membership studies based on optical/NIR photometry, optical spectroscopy, NIR variability, H$\alpha$ emission, X-ray emission and Gaia DR2/DR3 data (refer Section~\ref{sec:intro} for details). We also show the additional candidates identified in \citet{das2025} based on the deep optical (HSC) and NIR (UKIDSS) photometry using machine-learning (ML) techniques. Throughout this paper we consider these known members and candidate members as our cluster membership dataset, comprising a total of 827 sources (taken from Table 2 of \citet{das2025}). It is important to note that since this dataset is constructed from multiple studies, it may include some contaminants due to inherent biases in the underlying datasets and membership criteria. This is particularly relevant for the candidates identified in \citet{das2025}, however given the reported efficiency of their ML technique a clear majority of these candidates are likely to be genuine members.

Here we use the values reported in \citealt{das2023,das2025} for distance (917 $\pm$ 3 pc), mean age of the cluster (2 $\pm$ 1 Myr), and extinction A$_\mathrm{V}$ (1 $\pm$ 0.5 mag). The distance and its associated uncertainty were derived in \citet{das2023} as the weighted mean parallax of confirmed cluster members. We adopt the 10 Myr BHAC15 isochrone \citep{baraffe2015} reddened by an extinction of A$_\mathrm{V}$ = 1 mag and placed at a distance of 917 pc, as a guide to select candidate low-mass stars and brown dwarfs with r$_2$$>$20 mag (corresponding to mass $\sim$0.2 M$_\odot$). We use the 10 Myr isochrone as most of the previously known members are distributed to the right of this isochrone in the CMD. Targets for follow-up spectroscopy were selected from objects located to the right of this isochrone, the majority of which were candidate members identified by \citet{das2025}. In addition, some sources that were not classified as candidates in \citet{das2025} were also included as targets for follow-up spectroscopy to make optimal use of the observing time with the EMIR and SpeX instruments. We make use of the \citet{gordon2023} extinction relations and note that the BHAC15 isochrone for PanSTARRS1 (Vega) filter system were converted to HSC (AB) filter system using the transformation relations from \citet{gupta2021}. For clarity, we only show the 41 sources whose spectra could be retrieved after the data reduction, both with GTC-EMIR (12 sources) and IRTF-SpeX (30 sources), among which 1 source was observed with both instruments. One source (IC1396-20) lies close to the 10 Myr isochrone in this figure and appears bluer than the member sequence. When examined in the NIR, it is located well to the right of the 10 Myr isochrone along the member sequence, and is among the faintest J-band targets in our sample. This behaviour could be attributed to the presence of an edge-on circumstellar disk and/or due to magnetospheric accretion. Therefore, the photometric properties of this source should be interpreted with caution.

\subsection{Spectral type and extinction}
\label{sec:SpT}
\begin{figure*}
    \centering
    \includegraphics[scale=0.8]{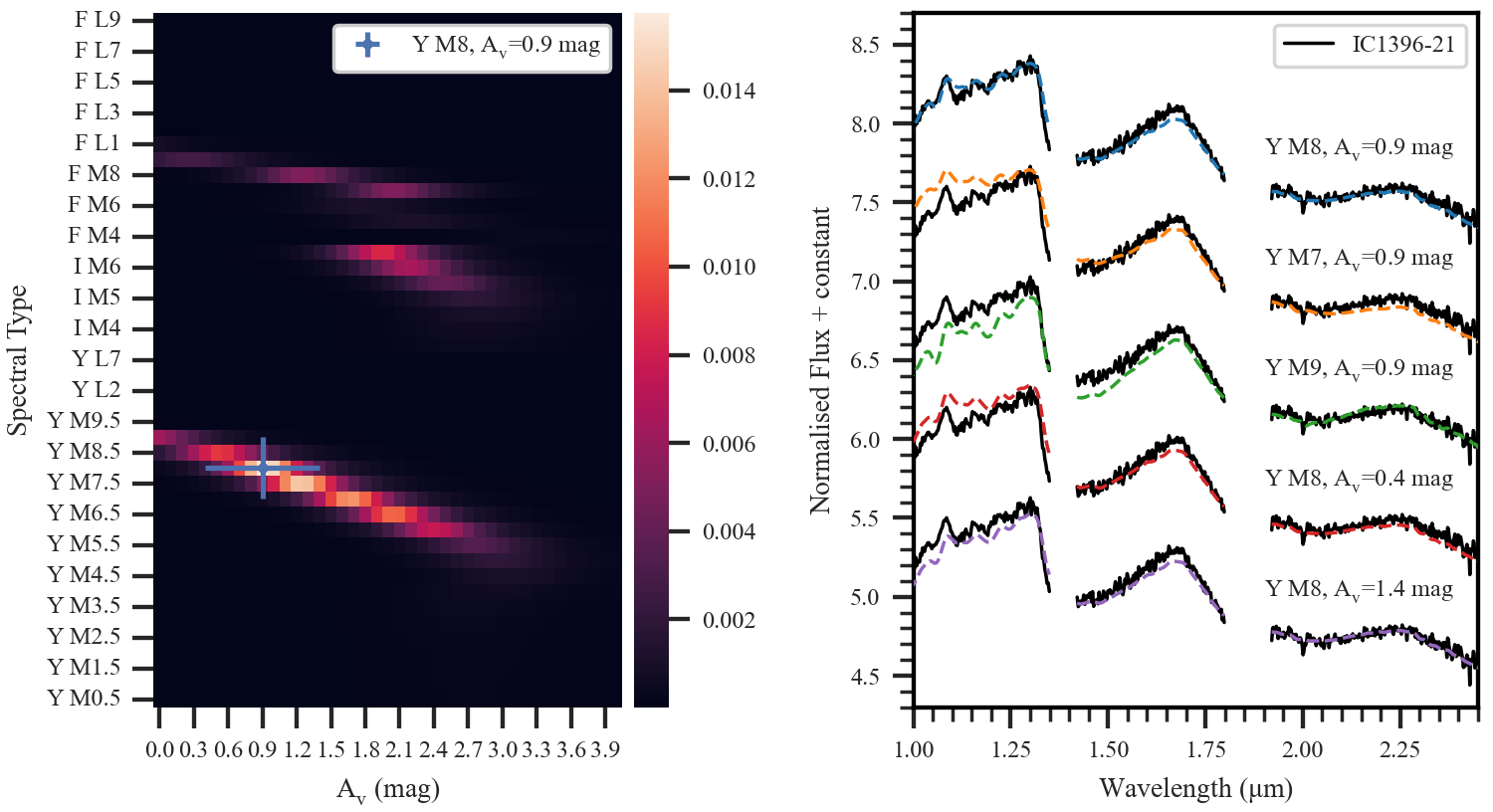}
    \caption{(left) A$_\mathrm{V}$ vs spectral type map of one of our targets (IC1396-21) observed with IRTF-SpeX. The order of templates in the y-axis has no significance and are only grouped according to their ages. The spectral types prefixed with ’Y’, 'I', and ’F’ indicate the young dwarfs, intermediate age dwarfs and old field dwarfs, respectively. The colour bar indicates the normalised 1/($\chi^2$)$^2$ value where the lowest $\chi^2$ corresponds to the brightest colour. The blue coloured marker denotes the best-fitting spectral type and A$_\mathrm{V}$ with an uncertainty of 1 spectral subtype and 0.5 mag in extinction (refer text for details). (right) SpeX spectra of the same object over plotted with the best-fitting reddened template (blue) re-sampled to match the target resolution. The templates with $\pm$1 subtype and $\pm$0.5 mag in extinction are shown in different colours for comparison.}
    \label{fig:spt_typing}
\end{figure*}

The spectral types of the sources were obtained by comparing the observed spectra with standard spectral templates. The procedure is similar to that detailed in \citet{damian2023a} and \citet{damian2025}. We used standard spectra of dwarfs of three age groups: 
\begin{itemize}
    \item young dwarfs ($<$5 Myr) of type M0-L0 (at intervals of 0.5 subclass), L2, L4 and L7 templates from \citet{luhman2017}.
    \item intermediate age dwarfs ($\sim$10 Myr) of type M3.5-M6.5 (at intervals of 0.5 subclass) from \citet{luhman2017}.
    \item old field dwarfs of type M4-L9 (for every 1 subclass) from the SpeX prism spectral library \citep{burgasser2014}.
\end{itemize}
To indicate the age of the templates, we use the prefix
’Y’ for young, 'I' for intermediate and ’F’ for old field dwarfs. The observed target spectra were compared with the reddened standard spectra using the equations below:
\begin{equation}
    \chi^2 = \frac{1}{N-2} \sum_{i=1}^{N} \frac{(O_i - E_i)^2}{{\sigma_i}^2} \quad \text{(for spectra observed with SpeX)} 
    \label{eq:reduced_chi_squared}
\end{equation}
\begin{equation}
    M = \frac{1}{N-2} \sum_{i=1}^{N}{(O_i - E_i)^2} \quad \text{(for spectra observed with EMIR)}
   \label{eq:reduced_M}
\end{equation}
\noindent Here \textit{N-2} is the degrees of freedom with temperature and extinction as the free parameters, \textit{O} is the observed target flux, \textit{E} is the template flux, and \textit{$\sigma$} is the uncertainty in the observed flux. We use the standard chi-square equation for all the targets observed with SpeX and a metric \textit{M} \citep{langeveld2024} for the targets observed with EMIR. We use this metric since we do not have the uncertainties in the observed fluxes from the EMIR data reduction pipeline. The standard spectra were reddened by varying the extinction A$_\mathrm{V}$ between 0 to 4 mag in steps of 0.1 mag and applying the \citet{gordon2023} extinction law. Figure~\ref{fig:spt_typing} shows as a sample the results of one of the objects observed with SpeX. We mask the prominent telluric absorption features and normalise the target spectra and the template spectra at 1.7 $\mu$m to accommodate the wavelength coverage of both SpeX and EMIR observations. We visually examine the parameter range close to the best fit templates and estimate the uncertainties as $\pm$1 spectral subtype and $\pm$0.5 mag in A$_V$.

We performed the spectral typing analysis for all the 41 targets. Nine sources were identified to best fit with old field dwarf spectra and 32 sources matched with young (Y) or intermediate (I) age standard spectra with age $<$10 Myr and spectral type between M3 and M9. One of the objects, IC1396-5 observed with both SpeX and EMIR best-fit with a young M5.5 and M6 template respectively. For two sources (IC1396-25 and IC1396-32), only the H-band spectra (1.5–1.8~$\mu$m) were used due to the very low signal-to-noise ratio at longer wavelengths. The spectral types of these 32 sources are presented in Table~\ref{tab:ic1396_objects} and their best-fitting template spectra are shown in Figures~\ref{fig:apen_spt_irtf} and~\ref{fig:apen_spt_gtc}. We consider these 32 objects with age $<$10 Myr as members of the cluster and exclude the remaining 9 sources as non-members. These members and non-members are highlighted in Figure~\ref{fig:cmd2}.

\subsection{Estimating effective temperature with model spectra}
\label{sec:teff_estimate}
We fit the spectra of our 32 sources identified as members in the previous section with the BT-Settl photospheric models \citep{allard2012} to estimate their effective temperatures (T$_\mathrm{eff}$). In Equations \ref{eq:reduced_chi_squared} and \ref{eq:reduced_M} we replace the degrees of freedom with \textit{N-4}, where the four free parameters are T$_\mathrm{eff}$, log g, A$_\mathrm{V}$ and the scaling factor. We use the models with temperatures ranging between 1500-4000 K (grid size 100 K) and log g between 3-5 (grid size 0.5). We vary A$_\mathrm{V}$ between 1-4 mag (grid size 0.1 mag) and the scaling factor between 10$^{-11}$ and 12$\times$10$^{-11}$ (grid size 10$^{-12}$). As in the previous analysis of estimating the spectral types, here too we mask the telluric absorption bands for the comparison. For IC1396-25 and IC1396-32, we use only the H-band spectra as discussed in the previous section and we note that the temperatures estimated for these two objects should be interpreted with caution, especially for IC1396-32, whose temperature is significantly lower than expected for its luminosity. Figures~\ref{fig:apen_teff_irtf} and ~\ref{fig:apen_teff_gtc} present the best-fitting results for all our targets observed with IRTF-SpeX and GTC-EMIR, respectively. The temperature, extinction, and log g estimated from this analysis for all the 32 sources are reported in Table~\ref{tab:ic1396_objects}. The extinctions estimated for most sources are consistent with those obtained in Section~\ref{sec:SpT}, with an average difference of $\sim$0.5 mag.

\begin{figure}
    \centering
    \includegraphics[width=0.9\linewidth]{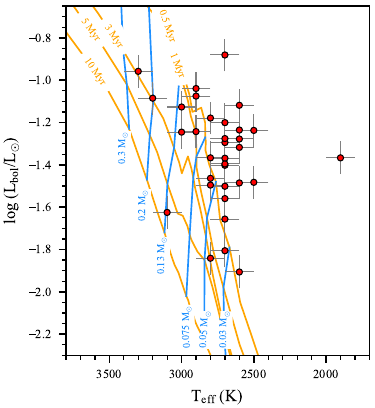}
    \caption{HR diagram of the 32 sources that were spectroscopically observed in this work and identified as members in Section~\ref{sec:SpT}. We show the BHAC15 \citep{baraffe2015} evolutionary tracks in cyan (masses 0.03, 0.05, 0.075, 0.13, 0.2 and 0.3 M$_\odot$) and isochrones in orange (ages 0.5, 1, 3, 5 and 10 Myr). }
    \label{fig:hrd}
\end{figure}

In Figure~\ref{fig:hrd} we show the Hertzsprung-Russell (HR) diagram for the 32 spectroscopically confirmed members. For the object IC1396-5 we have shown the temperatures (with corresponding luminosities) derived from both the EMIR and SpeX observations. The effective temperature of the members range between 1900-3300 K (see Table~\ref{tab:ic1396_objects}). To compute their bolometric luminosity (L$_\mathrm{bol}$), we use the J-band photometry from the UKIDSS (UKIRT Infrared Deep Sky Survey; \citealt{lawrence2007}) 11PLUS Galactic Plane Survey (GPS) catalogue. The bolometric magnitude is given by,
\begin{equation}
    M_{bol} = m_J - 5log(d) +5 - A_J + BC_J
\end{equation}
\noindent where \textit{m$_J$} corresponds to the J-band magnitude, \textit{d} is the distance (917$\pm$3 pc), and \textit{A$_{J}$} is obtained using the extinction values derived for each source in Section~\ref{sec:SpT} and the extinction relation from \citet{gordon2023}. The bolometric corrections (\textit{BC$_{J}$}) are assigned based on the spectral type, using the BC$_\mathrm{J_{MKO~young}}$ versus SpT relation from \citet{sanghi2023}. The bolometric luminosity is then calculated based on the bolometric magnitude using the equation shown below, where the solar bolometric magnitude, \textit{M$_{bol,\odot}$} is taken to be 4.73 mag.
\begin{equation}
    log(L_{bol}/L_\odot) =-\frac{(M_{bol}-M_{bol,\odot})}{2.5}
\end{equation}
All our targets are distributed to the right of the 10 Myr isochrone consistent with the colour and magnitude based candidate selection. One source (IC1396-32) appears anomalous, showing a relatively low temperature for its luminosity and spectral type. As noted above, the limited spectral coverage available for this object likely led to an underestimation of its temperature. 

\subsection{Spectral type versus effective temperature}
\label{sec:Spt_vs_teff}
\begin{figure}
    \centering
    \includegraphics[width=0.9\linewidth]{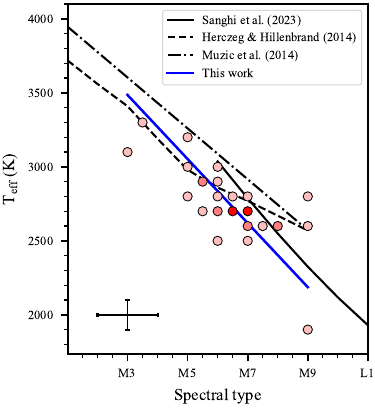}
    \caption{Comparison of spectral types and effective temperatures for the 32 spectroscopically confirmed members. Typical uncertainties of ±1 spectral subtype and ±100 K in temperature are indicated in the lower-left corner. The colour scale represents the number density of sources, with darker shades corresponding to higher densities. The blue line shows the best-fitting linear relation to the distribution of our members. We also show the relations from \citet{muzic2014}, \citet{herczeg2014}, and \citet{sanghi2023}.}
    \label{fig:spt_teff}
\end{figure}

In Figure~\ref{fig:spt_teff} we compare the distribution of our 32 targets with empirical temperature-spectral type relations from  \citet{herczeg2014}, \citet{muzic2014}, and \citet{sanghi2023} to assess how our temperature estimates compare with the standard literature relations. The relation from \citet{sanghi2023} is based on IR spectral types while the other two are based on optical data. Our data follow the \citet{herczeg2014} relation closely at earlier spectral types ($<$M6), while for later types ($>$M6) the trend is more consistent with the distribution observed in \citet{sanghi2023}, considering the uncertainties in temperature and spectral typing. However, in contrast, our objects have consistently lower effective temperatures than the relation reported by \citet{muzic2014} across the entire spectral range.

To further quantify the relation between spectral type and effective temperature in our sample while accounting for uncertainties in both parameters, we fit a linear relation using the Bayesian model of \citet{kelly2007}, implemented through the Python package \textit{linmix}\footnote{\url{https://github.com/jmeyers314/linmix?tab=readme-ov-file}}. The linear fit for our sample of 32 stars and brown dwarfs is of the form,
\begin{equation}
    T_{eff} = 4135\pm355 - (216\pm55) \times SpT
\end{equation}
where \textit{SpT} corresponds to the M subtype. We emphasize that the derived temperatures are model dependent and should therefore be interpreted in conjunction with, and within the limitations of, the adopted atmospheric models.

\section{Properties and substellar census of IC 1396}
\label{sec:discussion}
\subsection{Census of low-mass stars and brown dwarfs}
\label{sec:star_bd_census}
The deep optical photometric survey with HSC enables us to probe the substellar population in IC 1396 with a data completeness down to $\sim$0.03~M$_\odot$. We derive the masses of the sources in the membership catalogue previously compiled by \citet{das2025}, together with the new spectroscopically confirmed members presented in this work, using the HSC r$_2$-band photometry. In total there are 834 sources within the surveyed area of 22$\arcmin$ radius. Since spectral types are not available for all of them, we adopt photometry-based mass estimates that incorporate uncertainties in the photometry, cluster age, distance, and extinction using a Monte Carlo approach. We adopt a distance of 917$\pm$3 pc and an age of 2$\pm$1 Myr following \citet{das2023} and \citet{das2025}. For the extinction, we use the mean and standard deviation derived from the 32 spectroscopically confirmed members in this work (see Table~\ref{tab:ic1396_objects}), corresponding to A$_\mathrm{V}$ = 1.7$\pm$1.1 mag. For each cluster member, we generate 10,000 values for the four parameters--distance, age, extinction, and r$_2$-band photometry--from normal distributions centred on their mean values with standard deviations corresponding to the respective uncertainties. We then perform 10,000 iterations and in each iteration a random value for each of the four parameters is drawn from the previously generated distribution, and the (sub)stellar mass of the cluster member is estimated using the BHAC15 \citep{baraffe2015} isochrones. The final mass and its uncertainty for each source are taken as the median and standard deviation of the resulting mass distribution. 

Our spectroscopic follow-up of the photometric candidates allows us to place constraints on the number of possible missing members, especially brown dwarfs, in our membership catalogue. We adopt 0.075 M$_\odot$ as the upper mass limit for brown dwarfs and 0.03 M$_\odot$ as the lower limit, where the latter is set by the 90\% completeness limit of our photometry. Assuming an age of 2 Myr, an extinction of A$_\mathrm{V}$ = 1 mag, and a distance of 917 pc, these mass limits translate to $r_2$-band magnitudes of 22.3 mag and 24.5 mag, respectively. We use this photometric boundary as an approximate reference, while noting that individual sources near the boundary may not be robustly classified as stellar or substellar based on photometry alone. Within the brown dwarf mass range, we have 203 sources observed in the HSC survey that are present along the pre-main sequence branch, to the right of the 10 Myr isochrone (see Figure~\ref{fig:cmd2}). Among these, 66 sources are candidates or confirmed members identified either in previous photometric/astrometric membership studies (see Table 2 of \citealt{das2025}) or in our spectroscopic survey (refer Figure~\ref{fig:cmd2}). We obtained the spectra of 9 sources that were not identified as candidates in any of the previous works and based on our spectroscopic follow-up, 3 of them are non-members (see Section~\ref{sec:SpT}). This implies a membership confirmation rate of 66\% for this population of previously unclassified sources. Adopting this probability as representative for the remaining unclassified population, we estimate that 66\% of the 134 sources (203 – 66 – 3) are likely members. This implies that in addition to the sub-stellar members in our catalogue, we expect $\sim$88 more brown dwarfs (mass between 0.075 and 0.03 M$_\odot$) to be present in the surveyed region. 

In a similar manner when we consider the stellar mass range between 1 and 0.075 M$_\odot$, only one source that was previously unclassified in any of the membership studies was part of our spectroscopic follow-up. This target was determined to be a non-member based on our spectral analysis, hence it is likely that we do not miss any sources in this mass range.

\subsection{Star to brown dwarf ratio}
Theoretical models have proposed that external factors such as local stellar density and UV radiation fields can affect the fragmentation and accretion process involved in the formation of brown dwarfs (see Section~\ref{sec:intro}).  To observationally explore the effect of these factors on brown dwarf formation efficiency we calculate the star to brown dwarf ratio. This ratio measures the relative abundance of brown dwarfs with respect to stars in the region. 

For our analysis, we adopt a stellar mass range of 1-0.075 M$_\odot$ and a brown dwarf mass range of 0.075–0.03 M$_\odot$. The upper stellar mass limit is commonly adopted in studies of other star-forming regions, allowing direct comparison of their star to brown dwarf ratios. The lower brown dwarf limit of 0.03 M$_\odot$ is consistent with that typically used in the literature, and is also set by the completeness of our optical catalogue. To derive the star to brown dwarf ratio for IC 1396, we employ the Monte Carlo simulation for the 834 sources in our membership catalogue. Using the masses estimated in Section~\ref{sec:star_bd_census}, we run 10,000 iterations and for each iteration calculate the star to brown dwarf ratio by counting the number of sources within the stellar and substellar intervals defined above. To account for the probable brown dwarfs missing from our catalogue, we add a constant value of 88 (refer Section~\ref{sec:star_bd_census}) to the number of brown dwarfs in each iteration. 

\begin{figure}
    \centering   \includegraphics[width=\linewidth]{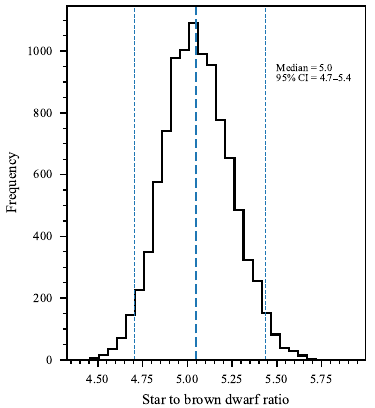}
    \caption{Histogram of star to brown dwarf ratio showing the results of the Monte Carlo simulation. The thick dashed line marks the median of the distribution and the two thin dashed lines show the 95\% confidence interval.}
    \label{fig:star_bd_ratio}
\end{figure}

The results of the Monte Carlo simulation are shown in Figure~\ref{fig:star_bd_ratio}. The median star to brown dwarf ratio is 5.0 and the 95\% confidence interval is between 4.7-5.4. This is however the upper limit to the star to brown dwarf ratio of IC 1396 as there may be brown dwarfs fainter than our survey limit, that is, below the lower mass limit of 0.03 M$_\odot$ considered here. Our estimate is lower than the ratio of 6$\pm$0.8 reported in \citet{gupta2024} who used the HSC photometry and machine learning methods to identify the members in the same region. The estimated ratio is derived from the available catalogue of candidates and confirmed members in the region and may therefore be refined with future spectroscopic confirmation of the complete sample. We emphasise that the quoted uncertainty reflects only the effects of the parameters included in the Monte Carlo analysis (distance, age, extinction, and photometric uncertainties). Additional systematic effects such as unresolved binarity, variability, accretion, circumstellar disks, and uncertainties in the adopted evolutionary models may further increase the uncertainty of the derived star to brown dwarf ratio.

One commonly used method to characterise the influence of environment on brown dwarf population, mainly the stellar density and feedback from massive stars, is to compare the star to brown dwarf ratio across different star-forming regions (\citealt{scholz2012,gupta2024,muzic2025}). Here we compare our results with those of eight young star-forming regions, namely Corona Australis (CrA), Cha I, Lupus 3, NGC 1333, NGC 2244, RCW 38, Tr 14, and RCW 36 (\citealt{scholz2012,muzic2015,muzic2017,muzic2025,dbdvale2026,rom2026}). These regions span a wide range of stellar densities and far-ultraviolet (FUV) radiation fields from massive OB stars, with stellar densities ranging from $\sim$10 to 4000 stars per pc$^2$ and log(Flux$_\mathrm{FUV}$) from -0.2 to 5.5 G$_\mathrm{o}$. To estimate the average stellar density and the incident FUV flux of IC 1396, we follow the same procedure used for the comparison regions, as outlined in \citet{muzic2025}. The details of this procedure are provided in Appendix~\ref{sec:appen_sdp}. It is important to note that all these regions have measured the star to brown dwarf ratio for the same stellar and sub-stellar mass interval as this study. 

\begin{figure}
    \centering
    \includegraphics[width=\linewidth]{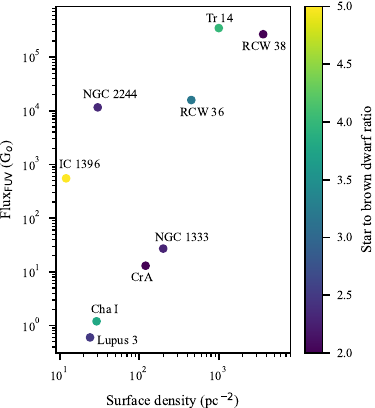}
    \caption{Star to brown dwarf ratio for star-forming regions listed in Table~\ref{tab:star_bd_ratio} as a function of the cluster stellar density and the median FUV flux. }
    \label{fig:fuv_density_sbd}
\end{figure}

Table~\ref{tab:star_bd_ratio} lists the ratios for the eight comparison regions and IC 1396, along with their stellar surface densities and FUV fluxes. For CrA, we adopt the star to brown dwarf ratio from \citet{muzic2025} derived using the BHAC15 isochrone, for which the underlying membership census builds on spectroscopically confirmed members compiled in \citet{luhman2022}. Similarly, the star to brown dwarf ratios for Lupus 3 and Cha I are adopted from \citet{muzic2025}. For Cha I the ratio in \citet{muzic2025} is based on the spectroscopic membership census presented in \citet{luhman2007}, \citet{luhman2008a} and \citet{luhman2008b}. For NGC 1333, we adopt the ratio from \citet{scholz2012}, which is based on spectroscopically confirmed members identified in that work and the earlier SONYC surveys (\citealt{scholz2009,scholz2012}). An updated membership census for NGC 1333 is presented in \citet{luhman2016}. However for the purpose of this work, we use the available star to brown dwarf ratio from \citet{scholz2012}, as re-estimating it is beyond the scope of this paper. We also include the regions NGC 2244, RCW 38, Tr 14 and RCW 36 adopting the star to brown dwarf ratios from \citet{muzic2019}, \citet{muzic2017}, \citet{rom2026}, and \citet{dbdvale2026}, respectively. For most regions in Table~\ref{tab:star_bd_ratio} the reported uncertainties represent the 95\% confidence interval, and the median values are used in Figure~\ref{fig:fuv_density_sbd}. The ranges of the ratios for Cha I and Lupus 3 are taken from \citet{muzic2025} and the average values used here were obtained through personal communication with the author. In general, the star to brown dwarf ratio ranges approximately between 2-5 and is consistent across the regions within the measured uncertainties. \citet{gupta2024} have done a similar analysis where they compare 15 star-forming regions in diverse environments and find the star to brown dwarf ratios to range between 2 and 6. They suggest that the brown dwarf fraction increases with stellar density but did not observe any significant influence of the FUV field strength.

\begin{table}
    \centering
    \caption{Star to brown dwarf ratios of different star-forming regions and their respective stellar densities and FUV fluxes.}
    \begin{tabular}{lccc}
    \toprule
    Region & Star to BD ratio & Stellar density & F$_\mathrm{FUV}$\\
    & & (pc$^{-2}$) & (G$_\mathrm{o}$) \\
    \midrule
        \vspace{0.2cm}
    IC 1396$^a$ & 5.0$^{+0.4}_{-0.3}$ & 12 & 551\\
        \vspace{0.2cm}
    Corona Australis (CrA)$^b$ & 2.0$^{+2.7}_{-1.0}$ & 120 & 13\\
        \vspace{0.2cm}
    Cha I$^b$ & 3.8$^{+1.0}_{-0.6}$ & 29 & 1.2\\
        \vspace{0.2cm}
    Lupus 3$^b$ & 2.5$^{+2.0}_{-0.4}$ &  24 & 0.6\\
        \vspace{0.2cm}
    NGC 1333$^c$ & 2.3$\pm$0.5 & 200 & 27\\
        \vspace{0.2cm}
    NGC 2244$^d$ & 2.34$\pm$0.25 & 33 & 1.16 $\times$ 10$^4$\\
        \vspace{0.2cm}
    RCW 38$^e$ & 2.0$\pm$0.6 &3620 & 2.66 $\times$ 10$^5$\\
        \vspace{0.2cm}
    Tr 14$^f$ & 4.0$^{+1.8}_{-1.2}$ & 1000 & 3.47 $\times$ 10$^5$\\
        \vspace{0.2cm}
    RCW 36$^g$ & 3.2$^{+2.1}_{-0.5}$ & 451& 1.58 $\times$ 10$^4$\\
    
    \bottomrule
    \end{tabular}
    \begin{flushleft} Note: The references for the star to BD ratios are (a) This work, (b) \citet{muzic2025}, (c) \citet{scholz2012}, (d) \citet{muzic2019}, (e) \citet{muzic2017}, (f) \citet{rom2026}, (g) \citet{dbdvale2026}. The stellar surface densities and FUV fluxes are taken from \citet{muzic2025} for CrA, Cha I, Lupus 3, NGC 1333, NGC 2244 and RCW 38, from \citet{rom2026} for Tr 14 and from \citet{dbdvale2026} for RCW 36. 
    \end{flushleft}
    \label{tab:star_bd_ratio}
\end{table}

Here we consider the average stellar surface density of the region and the median incident radiation from the OB stars in terms of FUV flux as a metric to compare the environments. Figure~\ref{fig:fuv_density_sbd} shows the star to brown ratio for IC 1396 and the other eight comparison regions, as a function of their respective average stellar density and incident FUV flux. IC 1396 has a stellar density comparable to the nearby regions like Cha I and Lupus 3 but has a much stronger radiation field. IC 1396 and NGC 2244 differ from the rest of the sample in that they have stellar densities comparable to those of nearby, low-density solar neighbourhood regions while being exposed to radiation fields several orders of magnitude stronger. However, the star to brown dwarf ratio of IC 1396 is higher than that of NGC 2244. Among the regions studied here, IC 1396 has the lowest brown dwarf fraction and the lowest source density. 

Considering the distribution of all the regions in the figure, at prima facie we do not see a definite trend to signify that brown dwarf formation has any dependence on the environmental conditions traced by stellar density and/or FUV radiation. Given the sample we have used, with our current knowledge it is difficult to establish the role of environment. Even if such a dependence exists, it may be challenging to detect considering the observational uncertainties and methodological differences. In particular, although broadly similar procedures are used to estimate stellar densities, radiation fields, and mass ranges across the nine regions, variations in the methods adopted to derive stellar masses may introduce additional systematic uncertainties. A larger, more uniformly analysed sample of star-forming regions is needed to probe deeper into influence of environment on brown dwarf formation.

\section{Summary}
\label{sec:summary}
We present the spectroscopic survey of low-mass stars and brown dwarfs in the young star-forming region IC 1396. This region provides an ideal laboratory beyond the solar neighbourhood to detect substellar objects in a feedback-driven environment, owing to its relatively low extinction caused by the presence of massive OB stars. Based on an optical photometric catalogue obtained with Subaru-HSC, we identified candidate low-mass stars and brown dwarfs in the region for subsequent spectroscopic characterisation. We conducted follow-up spectroscopy of these candidates using the SpeX instrument on NASA-IRTF and the EMIR multi-object spectrograph on GTC. This survey resulted in the spectroscopic confirmation of 32 members with spectral types between M3-M9 and effective temperatures between 1900-3300 K, all with estimated ages below 10 Myr.

Furthermore, we determined the star to brown dwarf ratio for IC 1396 to range between 4.7 and 5.4, with a median of 5.0 which is in agreement with the values derived for clusters in different nearby and distant star-forming environments. While this consistency suggests that the relative abundance of brown dwarfs may not strongly depend on local stellar environment, the limited sample size and associated uncertainties prevent a definitive conclusion. Overall, this work reports the first spectroscopic survey of brown dwarfs in IC 1396 and adds to the existing studies directed towards understanding their formation in feedback influenced regions.

\section*{Acknowledgements}
This research is based on data collected at Subaru Telescope with Hyper Suprime-Cam, which is operated by the National Astronomical Observatory of Japan. We are honoured and grateful for the opportunity of observing the Universe from Mauna Kea, which has the cultural, historical and natural significance in Hawaii. We are grateful to The East Asian Observatory which is supported by The National Astronomical Observatory of Japan; Academia Sinica Institute of Astronomy and Astrophysics; the Korea Astronomy and Space Science Institute; the Operation, Maintenance and Upgrading Fund for Astronomical Telescopes and Facility Instruments, budgeted from the Ministry of Finance (MOF) of China and administrated by the Chinese Academy of Sciences (CAS), as well as the National Key R\&D Program of China (no. 2017YFA0402700). The Pan-STARRS1 Surveys (PS1) and the PS1 public science archive have been made possible through contributions by the Institute for Astronomy, the University of Hawaii, the Pan-STARRS Project Office, the Max-Planck Society and its participating institutes, the Max Planck Institute for Astronomy, Heidelberg and the Max Planck Institute for Extraterrestrial Physics, Garching, The Johns Hopkins University, Durham University, the University of Edinburgh, the Queen's University Belfast, the Harvard-Smithsonian Center for Astrophysics, the Las Cumbres Observatory Global Telescope Network Incorporated, the National Central University of Taiwan, the Space Telescope Science Institute, the National Aeronautics and Space Administration under Grant No. NNX08AR22G issued through the Planetary Science Division of the NASA Science Mission Directorate, the National Science Foundation Grant No. AST–1238877, the University of Maryland, Eotvos Lorand University (ELTE), the Los Alamos National Laboratory, and the Gordon and Betty Moore Foundation. This work is partly based on data obtained with the instrument EMIR, built by a Consortium led by the Instituto de Astrofísica de Canarias. EMIR was funded by GRANTECAN and the National Plan of Astronomy and Astrophysics of the Spanish Government. Astronomer observing with the Infrared Telescope Facility, which is operated by the University of Hawaii under contract 80HQTR24DA010 with the National Aeronautics and Space Administration. 

BD and AS acknowledge support from the UKRI Science and Technology Facilities Council through grant ST/Y001419/1/. C. R-Z. acknowledge support from projects UNAM DGAPA PAPIIT IN 101723. We thank Koraljka {Mu{\v{z}}i{\'c}} for providing the average star to brown dwarf ratios for Cha I and Lupus 3 used in Figure~\ref{fig:fuv_density_sbd}.

\section*{Data Availability}
The data underlying this article will be shared on reasonable request to the corresponding author.



\bibliographystyle{mnras}
\bibliography{references} 




\appendix
\section{Newly identified members and their spectra}
\begin{figure*}
    \centering
    \includegraphics[width=\textwidth]{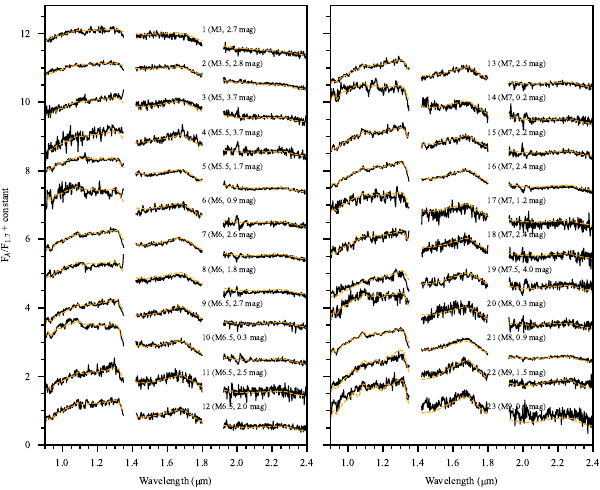}
    \caption{IRTF-SpeX spectra of all the 23 targets (black) normalised to the flux at 1.7 $\mu$m. The best-fitting template (orange) resampled to match the resolution of the SpeX spectra, normalised and reddened by the corresponding A$_\mathrm{V}$ for each source is over plotted. The suffix of the object ID (IC1396-*) is shown alongside each spectrum, together with the spectral type and A$_\mathrm{V}$ value.}
    \label{fig:apen_spt_irtf}
\end{figure*}

\begin{figure}
    \centering
    \includegraphics[width=\linewidth]{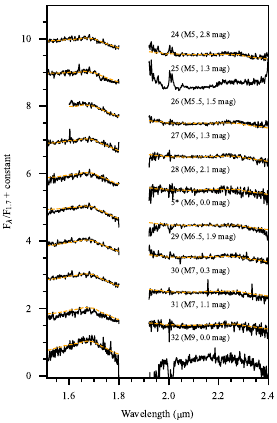}
    \caption{GTC-EMIR spectra of all the 10 targets (black) normalised to the flux at 1.7 $\mu$m. The best-fitting template (orange) resampled to match the resolution of the EMIR spectra, normalised and reddened by the corresponding A$_\mathrm{V}$ for each source is over plotted. The suffix of the object ID (IC1396-*) is shown alongside each spectrum, together with the spectral type and A$_\mathrm{V}$ value.}
    \label{fig:apen_spt_gtc}
\end{figure}

\begin{figure*}
    \centering
    \includegraphics[width=\textwidth]{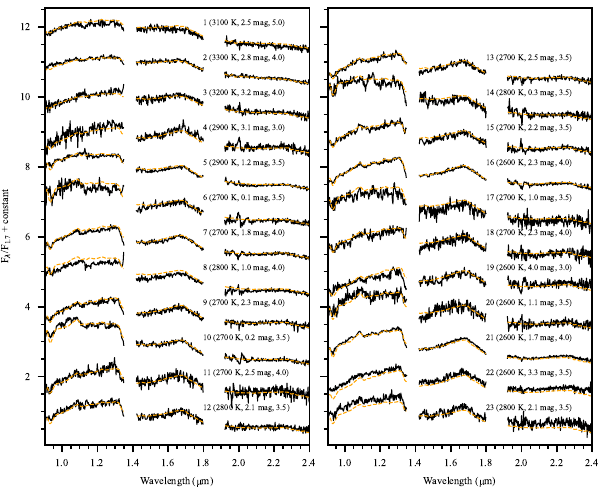}
    \caption{IRTF-SpeX spectra of all our 23 targets (black) normalised to the flux at 1.7 $\mu$m, same as in Figure~\ref{fig:apen_spt_irtf}. The best-fitting BT-Settl model spectra (orange) resampled to match the resolution of the object spectra is over plotted with T$_\mathrm{eff}$, log g and A$_\mathrm{V}$ as mentioned in the plot.}
    \label{fig:apen_teff_irtf}
\end{figure*}

\begin{figure}
    \centering
    \includegraphics[width=\linewidth]{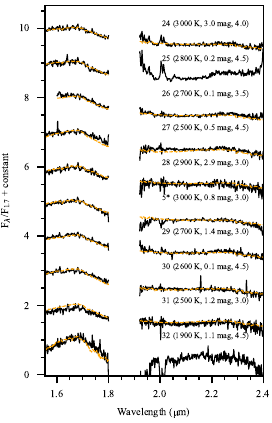}
    \caption{GTC-EMIR spectra of all our targets (black) normalised to the flux at 1.7 $\mu$m. The best-fitting BT-Settl model spectra (orange) resampled to match the resolution of the object spectra is over plotted with T$_\mathrm{eff}$, log g and A$_\mathrm{V}$ as mentioned in the plot.}
    \label{fig:apen_teff_gtc}
\end{figure}

Table~\ref{tab:ic1396_objects} presents the coordinates of all the 32 objects identified as members in this study along with their Subaru-HSC r$_2$, i$_2$, and Y-band photometry. The spectral type, effective temperature, log g, and extinction values of all the sources are also given in the Table (refer Sections~\ref{sec:SpT} and~\ref{sec:teff_estimate} for details). The mass of the objects and the associated uncertainties are also presented in the Table as discussed in Section~\ref{sec:star_bd_census}. In Figures~\ref{fig:apen_spt_irtf} and ~\ref{fig:apen_spt_gtc} we show the spectra of all the 32 objects over plotted with their best-fitting templates. Likewise Figure~\ref{fig:apen_teff_irtf} and ~\ref{fig:apen_teff_gtc} show the best-fitting atmospheric models for those newly identified members.  
\begin{table*}
    \centering
    \caption{Properties of the 32 spectroscopically observed members in this work. All of these sources best-fit with templates of age $<$10 Myr (see Section~\ref{sec:SpT} for details). Object IDs from IC1396-1 to IC1396-23 are targets surveyed with IRTF-SpeX and IC1396-24 to IC1396-32 are sources observed with GTC-EMIR (the sources are listed in increasing order of their spectral types). IC1396-5 was observed with both instruments and the '*' symbol indicates the observations with EMIR. The photometry obtained with HSC filters (r$_2$, i$_2$ and Y) are also provided.}
    \begin{tabular}{lccccccccccccc}
    \toprule
    Object ID & RA & Dec & r$_\mathrm{2}$ & e$_\mathrm{r_2}$ & i$_\mathrm{2}$ & e$_\mathrm{i_2}$ & Y & e$_\mathrm{Y}$ & SpT & T$_\mathrm{eff}$ & A$_\mathrm{V}$ & A$_\mathrm{V}$  & log g \\
    &  &  &  & &  & &  & &  & & template & model &\\
    & (deg) & (deg) & (mag) & (mag) & (mag) &(mag) & (mag) & (mag)& & (K) & (mag) & (mag) & \\
    \midrule
IC1396-1	&	324.2213	&	57.5187	&	23.6777	&	0.0090	&	20.0758	&	0.0044	&	18.4221	&	0.0010	&	M3	&	3100	&	2.7	&	2.5	&	5.0	\\
IC1396-2	&	324.3909	&	57.5380	&	19.7163	&	0.0002	&	18.0974	&	0.0002	&	16.9963	&	0.0002	&	M3.5	&	3300	&	2.8	&	2.8	&	4.0	\\
IC1396-3	&	325.1462	&	57.7821	&	23.3940	&	0.0040	&	19.8890	&	0.0004	&	18.2458	&	0.0004	&	M5	&	3200	&	3.7	&	3.2	&	4.0	\\
IC1396-4	&	324.1034	&	57.5140	&	22.7244	&	0.0034	&	20.1502	&	0.0011	&	18.1458	&	0.0006	&	M5.5	&	2900	&	3.7	&	3.1	&	3.0 \\
IC1396-5	&	324.8829	&	57.4855	&	23.3194	&	0.0060	&	19.2238	&	0.0005	&	17.6940	&	0.0005	&	M5.5	&	2900	&	1.7	&	1.2	&	3.5	\\
IC1396-6	&	324.3748	&	57.3592	&	22.4074	&	0.0021	&	19.3424	&	0.0006	&	17.7827	&	0.0005	&	M6	&	2700	&	0.9	&	0.1	&	3.5\\
IC1396-7	&	324.9999	&	57.7408	&	22.3893	&	0.0022	&	18.9356	&	0.0003	&	17.2581	&	0.0003	&	M6	&	2700	&	2.6	&	1.8	&	4.0 \\
IC1396-8	&	324.8855	&	57.4826	&	24.0775	&	0.0109	&	19.5673	&	0.0010	&	18.0757	&	0.0008	&	M6	&	2800	&	1.8	&	1.0	&	4.0	\\
IC1396-9	&	324.9855	&	57.7217	&	23.8052	&	0.0076	&	20.6740	&	0.0013	&	18.7131	&	0.0008	&	M6.5	&	2700	&	2.7	&	2.3	&	4.0	\\
IC1396-10	&	324.4811	&	57.4484	&	23.6859	&	0.0063	&	20.1123	&	0.0009	&	18.4018	&	0.0007	&	M6.5	&	2700	&	0.3	&	0.2	&	3.5	\\
IC1396-11	&	325.1879	&	57.4291	&	24.1497	&	0.0084	&	21.1480	&	0.0012	&	19.2468	&	0.0011	&	M6.5	&	2700	&	2.5	&	2.5	&	4.0	\\
IC1396-12	&	325.2095	&	57.7277	&	24.2167	&	0.0098	&	20.2997	&	0.0006	&	18.5659	&	0.0006	&	M6.5	&	2800	&	2.0	&	2.1	&	3.5	\\
IC1396-13	&	324.3343	&	57.7497	&	25.5545	&	0.0334	&	21.0804	&	0.0018	&	19.0064	&	0.0011	&	M7	&	2700	&	2.5	&	2.5	&	3.5	\\
IC1396-14	&	324.8800	&	57.5309	&	22.6828	&	0.0031	&	19.5043	&	0.0007	&	17.8327	&	0.0005	&	M7	&	2800	&	0.2	&	0.3	&	3.5\\
IC1396-15	&	325.3564	&	57.6057	&	23.1240	&	0.0033	&	20.3802	&	0.0006	&	18.4728	&	0.0005	&	M7	&	2700	&	2.2	&	2.2	&	3.5\\
IC1396-16	&	324.2571	&	57.3458	&	23.1623	&	0.0045	&	19.7356	&	0.0007	&	18.0597	&	0.0006	&	M7	&	2600	&	2.4	&	2.3	&	4.0\\
IC1396-17	&	325.1902	&	57.5677	&	24.2295	&	0.0088	&	20.8619	&	0.0009	&	19.1719	&	0.0009	&	M7	&	2700	&	1.2	&	1.0	&	3.5	\\
IC1396-18	&	324.3098	&	57.6429	&	23.6493	&	0.0055	&	20.5560	&	0.0011	&	18.6109	&	0.0007	&	M7	&	2700	&	2.4	&	2.3	&	4.0	\\
IC1396-19	&	324.3852	&	57.4135	&	24.9484	&	0.0205	&	21.7934	&	0.0038	&	19.6642	&	0.0021	&	M7.5	&	2600	&	4.0	&	4.0	&	3.0	\\
IC1396-20	&	325.0379	&	57.8191	&	23.6614	&	0.0058	&	21.4894	&	0.0021	&	19.4507	&	0.0011	&	M8	&	2600	&	0.3	&	1.1	&	3.5	\\
IC1396-21	&	324.5085	&	57.6752	&	23.2018	&	0.0061	&	19.7895	&	0.0006	&	17.9040	&	0.0006	&	M8	&	2600	&	0.9	&	1.7	&	4.0	\\
IC1396-22	&	324.4047	&	57.5531	&	23.8210	&	0.0083	&	19.2359	&	0.0005	&	17.7777	&	0.0005	&	M9	&	2600	&	1.5	&	3.3	&	3.5\\
IC1396-23	&	325.0511	&	57.6162	&	23.9158	&	0.0075	&	20.9590	&	0.0012	&	19.1269	&	0.0011	&	M9	&	2800	&	0.0	&	2.1	&	3.5	\\

IC1396-24	&	324.8019	&	57.3821	&	22.2558	&	0.0020	&	19.5597	&	0.0006	&	17.8677	&	0.0005	&	M5	&	3000	&	2.8	&	3.0	&	4.0	\\
IC1396-25	&	324.9158	&	57.4574	&	24.5932	&	0.0165	&	20.6519	&	0.0017	&	18.6401	&	0.0014	&	M5	&	2800	&	1.3	&	0.2	&	4.5\\
IC1396-26	&	324.8105	&	57.5304	&	23.3138	&	0.0056	&	19.6762	&	0.0011	&	18.0063	&	0.0008	&	M5.5	&	2700	&	1.5	&	0.1	&	3.5	\\
IC1396-27	&	324.8900	&	57.4710	&	21.8902	&	0.0016	&	19.5481	&	0.0006	&	17.8764	&	0.0006	&	M6	&	2500	&	1.3	&	0.5	&	4.5	\\
IC1396-28	&	324.6911	&	57.5515	&	22.3643	&	0.0020	&	19.5910	&	0.0006	&	18.0078	&	0.0006	&	M6	&	2900	&	2.1	&	2.9	&	3.0\\
IC1396-5*	&	324.8829	&	57.4855	&	23.3194	&	0.0060	&	19.2238	&	0.0005	&	17.6940	&	0.0005	&	M6	&	3000	&	0.0	&	0.8	&	3.0\\
IC1396-29	&	324.7736	&	57.3993	&	22.1672	&	0.0017	&	19.5065	&	0.0006	&	17.6950	&	0.0004	&	M6.5	&	2700	&	1.9	&	1.4	&	3.0\\
IC1396-30	&	324.8583	&	57.5155	&	21.6732	&	0.0013	&	19.2938	&	0.0007	&	17.7097	&	0.0004	&	M7	&	2600	&	0.3	&	0.1	&	4.5	\\
IC1396-31	&	324.7222	&	57.5303	&	23.0916	&	0.0044	&	20.4143	&	0.0015	&	18.4575	&	0.0008	&	 M7	&	2500	&	1.1	&	1.2	&	3.0	\\
IC1396-32	&	324.8397	&	57.4777	&	24.3649	&	0.0151	&	19.7470	&	0.0007	&	18.0265	&	0.0008	&	 M9	&	1900	&	0.0	&	1.1	&	4.5\\
\bottomrule
    \end{tabular}   
    \label{tab:ic1396_objects}
\end{table*}

\section{Estimating stellar surface density and FUV flux}
\label{sec:appen_sdp}
\begin{figure}
    \centering
    \includegraphics[width=\linewidth]{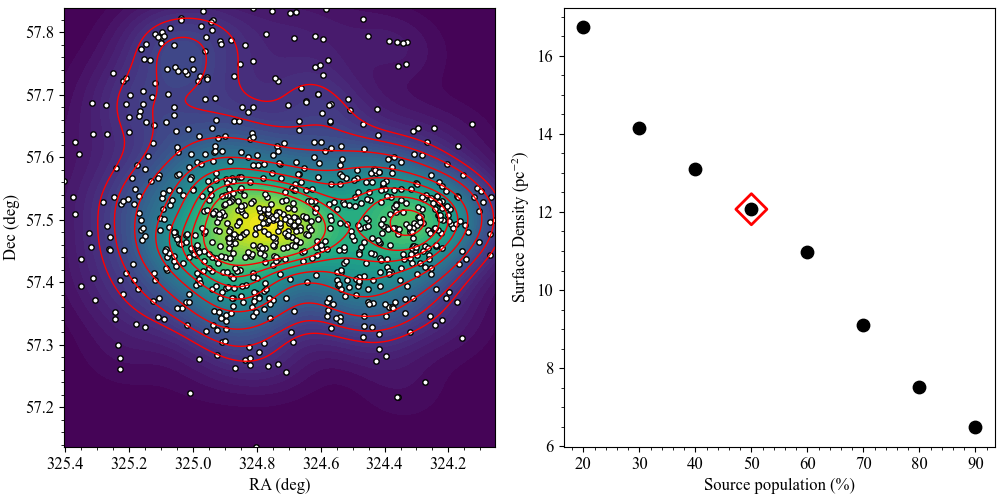}
    \caption{The left panel shows the spatial distribution of all 834 members (white dots) within the surveyed area of 22$\arcmin$ radius, overlaid on a kernel density map. The red contours delineate regions enclosing 20\% to 90\% of the population, in steps of 10\%, increasing outward. The right panel shows the corresponding surface density for each contour. The red diamond marks the surface density at the 50\% contour level which is considered as the average surface density of the region.}
    \label{fig:apen_sdp}
\end{figure}

We estimate the stellar surface density and FUV flux for IC 1396 following the same procedure as \citet{muzic2025}. In \citet{muzic2025}, only objects with spectral types earlier than M7 are considered but for IC 1396, since spectral types are not available for all sources, we instead use the full membership catalogue down to 0.03 M$_\odot$, which includes objects with spectral types as late as M9. As shown in Figure~\ref{fig:apen_sdp}, we trace the spatial distribution of the 834 stellar and substellar members in our surveyed area using a kernel density estimator. We plot density contours covering 20\% to 90\% of the population in increments of 10\% and calculate the corresponding surface density at each contour level. Following the reference point adopted by other studies for the comparison regions (refer Table~\ref{tab:star_bd_ratio}), we take the surface density corresponding to the 50\% contour level as the representative average value for the region. This way we find the stellar density for IC 1396 to be 12 stars per pc$^2$.

Likewise to derive the FUV radiation field of IC 1396, we follow a similar procedure as \citet{muzic2025}. First, we compile the list of massive stars, namely O and B-stars in the region taken from \citet{pelayo2023} and \citet{sicilia2005}. IC 1396 contains two O-stars (O5 and O9) and 16 B-stars (B1-B9) within the area of 22\arcmin radius. For each spectral subtype we obtain the corresponding effective temperature and bolometric luminosity from the extended table\footnote{\url{ http://www.pas.rochester.edu/~emamajek/EEM_dwarf_UBVIJHK_colors_Teff.txt}} of \citet{pecaut2013}. To measure the FUV luminosity of the massive stars, we scale their bolometric luminosity by a factor calculated using Equation 4 of \citet{kunitomo2021} for the objects temperature. The the FUV flux in units of G$_\mathrm{o}$ (G$_\mathrm{o}$ = 1.6x10$^{-3}$ ergs/s/cm$^2$) received by each member in the region is then the sum of the FUV luminosity from all the OB stars divided by the square of their distance from the member. As in \citet{muzic2025}, we assume that the distance along the line of sight is half of the projected separation in the plane of the sky. Finally, we take the median FUV flux incident on all the members in the region as the FUV field strength of the region. 

\bsp	
\label{lastpage}
\end{document}